# Where are the opportunities for growth in the professional services space?

24/03/2021

Edouard Ribes, CERNA Mines Paristech[1]

Abstract

The professional services industry (legal, accounting, consulting, architectural services...) employs an important share of the active population in mature countries. However, after decades of undisputed growth, the sector appears to be at a turning point in certain geographies. This article therefore proposes a simple framework to help diagnose where growth opportunities (if any) may lie.

When applied to the US economic context, the model indicates that at a macro-economic national level the sector should stall, which concurs with the trend observed over the past decade. However, it also highlights that a few industrial sectors (e.g. the US beverage industry) still offer pockets of growth for a variety of professional expertises. Replicating and fine-tuning those findings could be interesting for practitioners to steer their marketing and business development efforts. On the other hand, the quantitative framework presented in this study could pave the way for future research in the academic community.

## I - Introduction:

Over the past decades, the professional services industry has gained a considerable foothold in mature economies (US, EU, Japan etc....) (Empson, Muzio, Broschak, & Hinings, 2015). According to the data collected by the OECD (Organization for the Economic Cooperation & Development), the constituents of this sector now employ about 10% of developed countries' active population. Besides, the importance this sector has in mature societies keeps on increasing.

The professional services landscape is yet complex. It is made of a patchwork of expertises spanning from legal, accounting, auditing and management consulting know-hows to architectural and engineering capabilities (Maister, 2012). Although seemingly diverse, all those expertises are consumed by the society in the same very particular fashion: value is extracted as a client and an expert team up together to co-create a tailored solution (Løwendahl, 2005).

But knowledge, similarly to products and goods, is subject to a lifecycle. As the tenets of an expertise get commoditized, it becomes easier to service clients. The consequence is an increase in competitive pressure on professional services firms (referred to as PSFs in the rest of this article). This phenomenon has already been noticed in certain markets (e.g. the UK (Sako, 2006)) and PSFs are already adapting (Greenwood & Prakash, 2017). Evolution (not to say survival) recipes are well known and can be applied far beyond the confide of the professional services space. On one hand, firms can invest in technology to automate certain tasks and/or processes therefore gaining in productivity as well as developing new services (Frey & Osborne, 2017) (Acemoglu & Restrepo, 2019) (Acemoglu & Restrepo, 2020). On the other hand, PSFs can seek to relocate their resources as they grow in a cost-efficient manner (off/near-shoring, outsourcing etc...) (Jensen, Kletzer, Bernstein, & Feenstra, 2005).

---

[1] Contact: edouard-augustin.ribes@mines-paristech.fr

This type of race indexed on profitability and productivity metrics is a very old concept. All firms, independently of their industrial sector of origin, routinely engage in performance improvement actions (though with different level of success). This kind of hygiene activities is already in place in PSFs, a fact well documented in the academic literature (Sako & Tierney, 2005) (Pisani & Ricart, 2016) (Ribes, What does labor displacement mean for management consulting firms?, 2020). However, beyond notions of productivity and profitability, those firm hold a very specific relationship to growth (Maister, 2012). Their business model indeed consists in an expert selling a service and training apprentices who end up producing the deliverable with the client (Greenwood & Empson, 2003). The incentive for the expert is to grow revenue/client base to increase its personal earnings and to grow its pool of apprentice. On the other side, apprentices look forward to growing their expertise and to access new client of their own when their training is complete. Some might say that entering the professional services space requires a growth mindset, to the point that a stalling professional service practice may need to revise its incentive scheme and human resource paradigm (Ribes, 2020).

Striving for growth yet requires some focus (Coad, 2009). Opportunities are indeed made over time in a certain geographical and industrial context. This article therefore offers some advice in the form of economic modeling to help assess where such opportunities may lie. As such, this article should help advance the discussions on market research and client strategy dedicated to PSFs, a field that has, according to the review of (Skjølsvik, 2017), experienced a strong development over the past decade. Of course, most of the work associated to firm growth revolve around seizing the opportunity itself. Nevertheless, I hope that the simple considerations developed in this paper will prove useful for professional services experts as it will provide them with a replicable tool to orient their (personal) marketing efforts as well as for the academic community for it will incrementally build on the corpus of knowledge around firms' growth.

Structure-wise, this short article will start by a description of a toy model of firm growth (section II) and carry on with the analysis of its properties (section III). The associated theoretical results will then be applied to a public US dataset to highlight how the model can be used (section IV). A small discussion with concluding remarks (section V) will then be used to wrap the article and highlight potentially future avenues for debate.

## II – Modeling the professional services market:

The professional services market is made by the interactions of two types of stakeholders. On one hand, representatives of a firm in a certain industrial sector seeks a specific expertise to tackle challenges encountered by their enterprise. For instance, a manufacturing director could seek legal advice to craft a contract with a supplier in another country. On the other hand, an individual, who combines sectoral knowledge and a particular expertise, can position him/herself and some of his/her peers or staff to support the firm's representatives. For example, a lawyer specialized in international law who has a breath of experiences working in the manufacturing space could offer his/her services to the manufacturing director.

The sum of all those interactions lead to a market with two interdependent features. The first one is a level of price which reflects the performance of the matching process that binds a client and an expert. The second consists in the total number of experts who compete with one another to service the clients in their field. This section will therefore propose a simple model to assess the economic linkages between those two parties.

## Describing professional services consumers' behavior:

At a point in time $t$, an industrial sector is made of $f(t,r)$ firms generating a revenue $r$. Each of them is potentially interested by contracting the service of an expert. This service comes at a price $p$ and generates a benefit proportional to the firm size (up to a factor $v$). Therefore, the only firms who actually consume the service are the ones yielding a profitable outcome from the associated contract (i.e. $r \geq p/v$). The demand $D$ for a set expertise can therefore be expressed as:

$$D(t,p) = \int_{\frac{p}{v}}^{+\infty} f(t,r) dr$$

In any sector, firms are subject to a lifecycle (Penrose, 1952): they are born, they grow, they die. From a modeling standpoint, existing firms in a sector have a survival rate of $\mu$ and if they survive, they grow (Gibrat, 1931) at a rate $\psi$ (i.e. $dr = \psi.r.dt$). Besides, at every timestep, a proportion $\alpha$ of new firms comes to life with a revenue $r_m$ (i.e. $f(t,r_m) = \alpha. \int f(t,r)dr$). As a result, the evolution of the field is given by:

$$\forall r \geq r_m;\ \partial_t f + \partial_r(\psi.r.f) = -\mu.f$$

**Property 1.** In a sector where firms grow at a rate $\psi$, have a survival probability of $\mu$ and where the portion of new firms ($\alpha$) created at every time step is constant, the distribution of firms (i.e. $f$) grows exponentially from a longitudinal standpoint (i.e. with respect to time) and follows a power law from a sectional point of view (i.e. with respect to their revenue $r$):

$$\forall t, \forall r \geq r_m, f(t,r) = f_0.e^{(\alpha.\psi - \mu).t}.\left(\frac{r}{r_m}\right)^{-(1+\alpha)}$$

## Depicting the production model of professional services' providers:

On the other hand, the dynamics of professional services firms follow a slightly different pattern. If they have a lifecycle with a survival probability $\rho$ and a growth rate[2] $\phi$, the rate at which new PSFs enter the marketplace cannot be assumed constant. The entry of new firms is indeed conditioned by the pace at which its client pool grows. Calling $g(t,s)$ the number of PSFs at time t and $h(t)$ the number of new firms entering the market with $s_m$ experts (i.e. $g(t,s_m) = h(t)$), the structural assumptions of the model (i.e. $\partial_t g + \partial_s(\phi.s.g) = -\rho.g$) impose that the distribution of PSFs follows:

$$\forall s \geq s_m;\ g(t,s) = h\left(t - \frac{1}{\phi}.\ln\left(\frac{s}{s_m}\right)\right).e^{-\rho.\left(t - \frac{1}{\phi}.\ln\left(\frac{s}{s_m}\right)\right)}$$

This structure can be used to infer how many clients the sector can supply services to ($S(t)$). If the delivery of a service offering requires $n$ experts, the supply of service follows:

$$\forall t > 0;\ S(t) = \frac{1}{n}\int s.g(t,s).ds = \frac{s_m^2.\phi}{n}.e^{2\phi.t}\int_0^t e^{-(2.\phi+\rho).x}.h(x).dx$$

Beyond simple growth dynamics, professional services firms are also subject to efficiencies of scale. As they evolve, the production costs $c(s)$ of their services decreases (i.e. $\partial_s c < 0$). This can for instance be achieved through technological investments as well as by choosing an adequate set of production locations. In this model, production costs are assumed to decrease according to a power law (i.e. $dc = -\frac{\theta}{s}.c.ds \leftrightarrow c(s) = C_m.\left(\frac{s}{s_m}\right)^{-\theta}$). As a result of this production structure, small firms are not necessarily viable / profitable if prices are too low (i.e. if $p - n.c(s) < 0$). This recoups the general theory of growth and notably the idea that, in each industrial sector, firms have a minimal efficient size (Evans, 1987).

---
[2] This represents the ability those firms have to train new experts (i.e. $ds = \phi.s.dt$).

**Property 2.** In a market where professional services come at a price $p$, for services providers to survive, they must have a size $s$ such that:

$$s_m . \left(\frac{p}{n.C_m}\right)^{-\frac{1}{\theta}} < s$$

**Corollary 1.** In a market where PSFs enter the market with a size $s_m$, there can be no entry if prices are too low (i.e. $p < n.C_m$).

# III - Matching industrial consumer's behavior and provider's production abilities:

One core economic principle is that, at equilibrium, the price of a professional service should be such that its demand and supply levels are equal (i.e. $D(t,p) = S(t)$). This equilibrium can yet take two forms. The structure highlighted in the previous section indeed shows that if the equilibrium price is high enough (i.e. $p \geq n.C_m$), the professional services sector expands as new firms enter the market (i.e. $h(t) > 0$). On the other hand, a low equilibrium price leads to a contraction of the professional services space. This section will therefore investigate those two equilibrium states and showcase their properties.

## What are the conditions leading to the growth of a professional service?

For the professional services space to grow (i.e. $h(t) > 0$), prices must be high enough to ensure that new entrants can operate profitably (i.e. $p \geq n.C_m$). On the other hand, for the maximum of industrial firms to benefit from professional services[3] (i.e. $\max_p D(t,p)$), prices must be as low as possible. This leads to the following lemma.

**Lemma 1.** In a growing professional services market, the objective of customers' welfare maximization yields a level of price equal to:

$$p = n.C_m$$

Now, assuming that the market always adjusts itself to follow the evolution of its customers in a context of growth (i.e. $\forall t > 0, p = n.C_m, \partial_t D(t, n.C_m) = \partial_t S(t)$), the structural assumptions highlighted in the previous section yields that the flow of new entrants in the professional services space obeys:

$$h(t) = e^{(\alpha.\psi - \mu + \rho).t} n.(\alpha.\psi - \mu - 2.\phi).\frac{f_0}{s_m^2.\phi}.\frac{1}{\alpha}\left(\frac{n.C_m}{v.r_m}\right)^{-(\alpha)}$$

**Property 3.** In a context of growth, the professional services sector expands at an exponential rate and mimics the evolution of its clients (i.e. both the demand ($D(t, n.C_m)$) and the supply ($S(t)$) of professional services grow at a speed equal to $\alpha.\psi - \mu$). This type of evolution is yet only possible if the speed at which new professional experts are trained (i.e. $\phi$) is lower than the speed at which their customers' pool grows:

$$\alpha.\psi - \mu > 2.\phi$$

---

[3] Note that seeking to maximize demand is (in this case) equivalent to maximizing the welfare of an industrial sector. Given that the use of professional services enhances the profitability of an industrial sector's constituents, the overall profitability/ well-fare of the sector increases with the number of firms accessing professional experts.

In a context of growth, professional services firms are always profitable. However, efficiencies of scale trigger a heterogeneity in terms of performance. Larger firms are indeed more profitable than their smaller counterparts and the structural assumptions of the model yield that PSFs profitability ($\Pi$) obeys a power law (i.e. $\Pi(s) = s.n.C_m(1 - \left(\frac{s_m}{s}\right)^\theta)$). Operating a PSF in such an environment therefore generates a surplus (i.e. $\forall s > s_m, \Pi(s) > 0$) and it would be interesting to understand how this excess is used and redistributed. If this is out of scope of this article, this could potentially represent an interesting avenue for future discussions.

### What is happening if a professional expertise can be easily acquired?

The previous sub-section has shown that for a professional services market to grow, the pace at which experts can be trained must be below the pace at which the pool of clients grows (c.f. property 3). But when professional services practices' training capacity is more important than this threshold, there is an excess in professional experts in the market. In this kind of situation, prices must go down so that the market can self-regulate.

When prices start to decrease (i.e. $p < n.C_m$), the structural assumptions of the model lead to a certain degree of consolidation in the professional services field. Pricing conditions indeed deter entry since new firms of size $s_m$ cannot operate profitably. Besides, only the firms large enough to benefit from sizeable efficiencies of scale can survive (i.e. $s_m.\left(\frac{n.C_m}{p}\right)^{\frac{1}{\theta}} < s$).

**Property 4.** In a context where professional services firms start to consolidate (i.e. new entrants can no longer get a foothold), the price of an expertise decrease at an exponential rate:

$$p = n.C_m.e^{\frac{\theta}{2-\alpha.\theta}.(2\phi - \alpha.\psi + \mu).t}$$

Note that for price to be pressured down, the pace at which clients' revenue grows must be inferior to a threshold dependent in the speed at which PSFs can benefit from efficiencies of scale (i.e. $\frac{2}{\theta} < \alpha$). Competitive prices pressures create additional exits (on top of the ones caused by random managerial failures represented by the parameter $\rho$) as the minimal efficient size of a professional services firm (i.e. $s_m.e^{\frac{-1}{2-\alpha.\theta}.(2\phi - \alpha.\psi + \mu).t}$) also increases at an exponential rate.

## IV – What can we learn when we apply the model to the current US context?

The model developed through section II and III can be used to rapidly assess where opportunities of growth for PSFs lie in a certain geography. Property (3) indeed shows that by simply comparing the growth at which an industrial sector grows (i.e. $\alpha.\psi$) and is prone to managerial failure (i.e. $\mu$) with the speed at which experts can be trained (i.e. $\phi$), it is easy to map where entry is likely to be sustainable. This will be exemplified in the next sub-sections by applying the model to public data maintained by the US Census bureau.

### How to leverage publicly available information?

To calibrate the proposed model, data maintained by the US administration is used. Its dataset called "Business Dynamics Statistics" [BDS] contains publicly available data ranging from 1978 up to 2019. It displays the number of existing firms, employment figures, the number of exiting and entering firms in every industrial sector of the US economy[4]. The dataset presents 260+ unique sectors[5] identified according to the North American classification system (NAICS). Let us call $N_i(t)$ the number of existing firms in a sector $i$, $E_i(t)$ the employment in the sector at time $t$ and $N_i^{new}(t)$ (resp. $N_i^{exit}(t)$) the number of new (resp. exiting) firms in a sector reported in the dataset. Those datapoints can be used to derive some estimates of the key parameters underpinning the model over a period $T$.

From a consumer's dynamics point of view (i.e. all sectors excluding professional services ones[6]):

- the probability of a failure ($\mu_i$) in a sector $i$ can be approximated as $\mu_i = \frac{1}{T}\sum_T \frac{N_i^{exit}(t)}{N_i(t)}$
- the rate at which new firms enter the US market ($\alpha_i$) in a sector $i$ can be estimated as $\alpha_i = \frac{1}{T}\sum_T \frac{N_i^{entry}(t)}{N(t)}$
- the pace at which the revenue of a firm grows in a sector $i$ can be quickly valued under the assumption that labor productivity in the sector has remained unchanged[7] as $\psi_i = \frac{1}{T} \cdot \sum_T \left( \frac{E_i(t+1)-E_i(t)}{E_i(t)} \cdot \frac{N_i^{entry}(t)}{N(t)} + \frac{N_i^{exit}(t)}{N_i^{entry}(t)} \right)$

On the service providers' side (i.e. the professional services sector):

- the rate at which practices fail ($\rho_i$) in a sector $i$ can be approximated as $\rho_i = \frac{1}{T}\sum_T \frac{N_i^{exit}(t)}{N_i(t)}$
- the training speed of new experts ($\phi_i$) in a professional sector $i$ can be approached as $\phi_i = \frac{1}{T} \cdot \sum_T \left( \frac{E_i(t+1)-E_i(t)}{E_i(t)} \cdot \frac{N_i^{entry}(t)}{N(t)} + \frac{N_i^{exit}(t)}{N_i^{entry}(t)} \right)$

In the context of this article, estimates where drawn over a period of $T$ equal to 10 years starting from 2008. However, if the proposed estimates are very easy to obtain, they are rather crude. Customers' revenue growth estimates could indeed be improved by accounting for the pace at which labor productivity evolves over time. If some longitudinal information is made available by the US administration, it is only comprehensive at a high level (i.e. groups of NAICS sectors) and was not included in this analysis.

Professional services practices' dynamics are also subject to caution. Rates of failures ($\rho_i$) are likely to be overestimated with the proposed methodology in a market which is ongoing consolidation. In this set-up, several firms exit not because of managerial failure but because they are no longer competitive. It is however difficult to estimate the proportion of PSFs exiting the market because of productivity pressures with the available data. Additionally, this could create a bias in the estimation of experts training speed. Note that investing in some research towards the calibration methodology of population models towards standard economic datasets could represent an area of future research that would prove valuable in the long run.

### Key insights from the US professional services landscape:

By leveraging the BDS dataset, it is possible to estimate the speed at which professional services firms can train new experts (i.e. $\phi$). Results, displayed in table 1, show an important disparity across capabilities. For instance, the hardest skills to absorb appear to be in the accounting industry, whilst management consulting expertise appear to be easy to pick up. Calibration findings indeed yield that over a period of 10 years, 10 accounting experts are required to train a new peer, whilst only 3 management consulting experts are necessary to yield a similar outcome in their own field.

---

[4] https://www.census.gov/programs-surveys/bds.html
[5] Note that in this analysis, the 4th level of granularity in the NAICS classification was used.
[6] Professional services sectors are identified through the NAICS code 5411,5412,5413,5415,5416,5417,5418 & 5419.
[7] This assumption, although crude, implies that the rate at which the revenue of a sector grows over time (i.e. $\alpha.\psi - \mu$ in the model) is similar to the rate at which employment grows.

*Table 1 - Professional services firms survival probability and training speed.*

| NAICS Sector code | NAICS Sector description | Observed exit rate ($\rho$) | Estimated training speed ($\phi$) |
|---|---|---|---|
| 5411 | Legal services | 4,15% | 0,79% |
| 5412 | Accounting, tax preparation, bookkeeping, and payroll services | 5,32% | 0,45% |
| 5413 | Architectural, engineering, and related services | 12,91% | 1,35% |
| 5414 | Specialized design services | 6,42% | 1,52% |
| 5415 | Computer systems design and related services | 9,19% | 2,56% |
| 5416 | Management, scientific, and technical consulting services | 9,96% | 3,02% |
| 5417 | Scientific research and development services | 7,88% | 1,07% |
| 5418 | Advertising, public relations, and related services | 9,03% | 1,24% |
| 5419 | Other professional, scientific, and technical services | 9,58% | 0,91% |

Comparing the estimates drawn on professional services firms (i.e. $\phi$) to the entire US economy (i.e. $\alpha.\psi - \mu$), property (3) suggests that all professional services subsectors are in a phase of consolidation. The pace at which experts can be trained in those fields exceeds the pace at which the overall economy evolves (for the overall US economy $\alpha.\psi - \mu = 0,76\%$). At this macro-economic level, the output of the model appears in line with empirical observations. The numbers of firms in the professional services space have plateaued over the past 10 years within the US economy after having experienced a very strong growth since the late 70s (see figure 1)[8].

---

[8] Note that the management consulting space appears to be the exception here. This could potentially be explained by the development of entirely new branches of expertise in those fields over the past 10 years (e.g. artificial intelligence…), whilst the core skills of other fields have remained relatively static. In this case, it would probably be of interest to use a more advance industrial classification system.

*Figure 1 - Evolution of the number of PSFs over the past 4 decades in the US.*

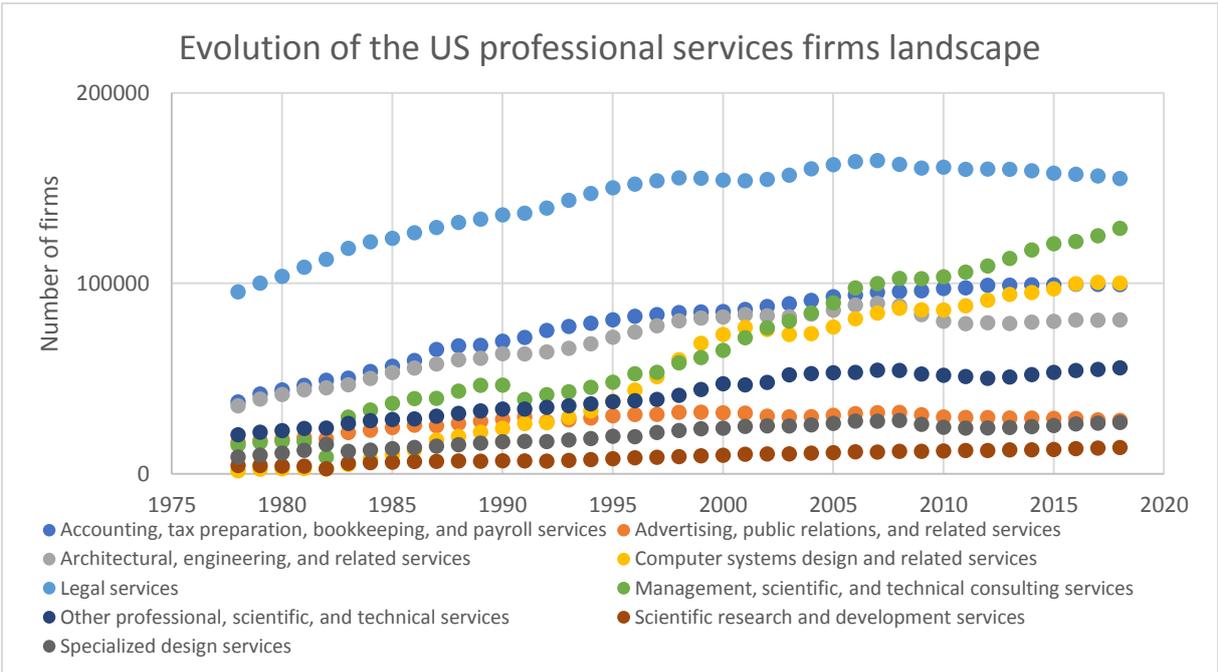

But this rough aggregated picture hides certain disparities. The development of several industrial sectors continues to spur the birth of new professional services practices. Examples of such sectors include the delivery industry, the beverage industry, the pipeline industry etc…. This can in turn explain why establishment creation rates continue to be high in those sectors despite an overall trend towards consolidation. The heatmap provided in table 2 provides such an illustration by showing where opportunities may exist for professional services experts across the top 10 fastest growing sectors in the US economy. In the heatmap, red is associated to a consolidation of the expertise applied to the industrial sector, whilst green highlights opportunities for growth.

*Table 2 - Opportunities for growth amongst professional services' expertises.*

| | Legal services | Accounting, tax preparation, bookkeeping, and payroll services | Architectural, engineering, and related services | Specialized design services | Computer systems design and related services | Management, scientific, and technical consulting services | Scientific research and development services | Advertising, public relations, and related services | Other professional, scientific, and technical services |
|---|---|---|---|---|---|---|---|---|---|
| Local messengers and local delivery | | | | | 🟥 | 🟥 | | | |
| Beverage manufacturing | | | | 🟥 | 🟥 | 🟥 | | | |
| Pipeline transportation of crude oil | | | | 🟥 | 🟥 | 🟥 | | | |
| Motor vehicle body and trailer manufacturing | | | | 🟥 | 🟥 | 🟥 | | | |

| Industry | | | | | | | | | |
|---|---|---|---|---|---|---|---|---|---|
| Support activities for rail transportation | | | | | | | | | |
| Motor vehicle manufacturing | | | | | | | | | |
| Electronic shopping and mail-order houses | | | | | | | | | |
| Management, scientific, and technical consulting services | | | | | | | | | |
| Amusement parks and arcades | | | | | | | | | |
| Other telecommunications | | | | | | | | | |
| Employment services | | | | | | | | | |

# V - Concluding remarks:

The model proposed in this article can be used to offer a simple view of where business development opportunities may exist for professional services experts. Its core consists in a comparison of the pace at which the revenue generated by an industrial sector grows against the speed at which professional experts can be trained. Its application to the US economy shows that a macro-economic level, all professional services spaces are ongoing a consolidation but that some industrial sectors still offer growth opportunities at a national level. The current calibration exercise notably highlights that the local delivery industry and the beverage manufacturing one are interesting prospecting grounds. From a practitioner standpoint, it could now be interesting to see how the model fares in terms of services prices prediction [should a suitable dataset including labor costs details be available] and potentially stress test it a more local level (e.g. US state or metropolitan area). From an academic point of view, it could also be of interest to investigate the nuances arising when more complex aspects of the lifecycle of firms are considered (i.e. change in growth speed and survival probability over time).

Sako, M., & Tierney, A. (2005). Sustainability of business service outsourcing: The case of human resource outsourcing (HRO). *Advanced Institute of Management Research Paper*.

Skjølsvik, T. P. (2017). Strategic management of professional service firms: Reviewing ABS journals and identifying key research themes. . *Journal of Professions and Organization*, 203-239.
11